\documentclass[journal]{IEEEtran}
\usepackage{threeparttable}
\usepackage{cite}
\ifCLASSINFOpdf
  \usepackage[pdftex]{graphicx}
  \graphicspath{{../pdf/}{../jpeg/}}
  \DeclareGraphicsExtensions{.pdf,.jpeg,.png}
\else
  \usepackage[dvips]{graphicx}
  \graphicspath{{../eps/}}
  \DeclareGraphicsExtensions{.eps}
\fi
\usepackage{amsmath}
\usepackage{amssymb}
\usepackage{algorithmic}
\usepackage{array}
\ifCLASSOPTIONcompsoc
  \usepackage[caption=false,font=normalsize,labelfont=sf,textfont=sf]{subfig}
\else
  \usepackage[caption=false,font=footnotesize]{subfig}
\fi
\usepackage{fixltx2e}
\usepackage{stfloats}
\usepackage{setspace}
\usepackage{mathrsfs}
\usepackage{url}
\begin{document}
	\title{ Ultraviolet Scattering Communication Using Subcarrier Intensity Modulation over Atmospheric Turbulence Channels}
	\author{Zanqiu~Shen,
		Jianshe~Ma,
		Tianfeng~Wu,
		Tao~Shan,
		Yupeng~Chen,
		and~Ping~Su
	}
\markboth{ }%
{Shell \MakeLowercase{\textit{et al.}}: Bare Demo of IEEEtran.cls for IEEE Journals}
\maketitle
\begin{abstract}
A closed-form non-line-of-sight (NLOS) turbulence-induced fluctuation model is derived for ultraviolet scattering communication (USC), which models the received irradiance fluctuation by Meijer G function. Based on this model, we investigate the error rates of the USC system in NLOS case using different modulation techniques. Closed-form error rate results are derived by integration of Meijer G function. Inspired by the decomposition of different turbulence parameters, we use a series expansion of hypergeometric function and obtain the error rate expressions by the sum of four infinite series. The numerical results show that our error rate results are accurate in NLOS case. We also study the relationship between the turbulence influence and NLOS transceiver configurations. The numerical results show that when two-LOS link formulates the same distance, the turbulence influence is the strongest for long ranges and the weakest for short ranges.
	
\end{abstract}
\section{Introduction}
\IEEEPARstart{U}{ltraviolet} scattering communication is proposed to achieve a NLOS optical communcation link, which has been studied  for decades by both theoretical analysis and experimental validation \cite{vavoulas2019survey}. Although NLOS channel models with atmospheric conditions and geometry parameters have been extensively studied, the NLOS turbulence induced fading model is still demanding, especially for studying the relationship between the turbulence influence and transceiver configurations.

To solve the challenges of NLOS turbulence channel modeling, an analytical model is proposed by assuming single scattering and weak turbulence conditions, which seperates the NLOS link into two-LOS link and models each path as lognormal irradiance fluctuation \cite{ding2011turbulence}. Base on \cite{ding2011turbulence}, another turbulence model is proposed by treating two-LOS link as independent links and the NLOS performance is evaluated for pulse position modulation \cite{liu2015performance}. Since the aforementioned models can only be suitable for weak turbulence conditions, Wang et al. proposed a Monte-Carlo simulation framework that models each LOS link as Gamma-Gamma distribution and study the turbulence influence in NLOS case \cite{wang2013characteristics}. However, the authors did not derive the closed-form results for NLOS turbulence channel. Thereafter, Arya et al. derived a closed-form expression of the received irradiance fluctuation based on Gamma-Gamma turbulence model assuming normalized irradiance for two-LOS link \cite{arya2019amplify}. However, the authors did not investigate the NLOS turbulence characterics in terms of different transceiver elevation angles. 

After the turbulence channel model is established, the communication performance degradation due to turbulence-induced fluctuationc needs to be settled in USC system, which has been extensively stduied in OWC system using ON-OFF keying (OOK) modulation. For USC systems, OOK modulation and direct detection has been used for experiment owing to its simplicity but with limited data rates \cite{noshad2013nlos}. To improve the communication performance of USC system, Noshad et al. used M-ary spectral-amplitude-coding to support high data rates and longer distance by sacrificing the spectral efficiency for the same performance compared to OOK modulation \cite{noshad2013nlos}. However, the M-ary spectral-amplitude-coding technique requires complex tranceiver design and the authors did not study the turbulence induced fading for this modulation. As another modulation technique, subcarrier intensity modulation (SIM) provides better communication performances compared to OOK modulation. Therefore, Popoola et al. studied the error rate performance using binary phase shift keying (BPSK) modulation for weak to strong turbulence conditions and saturation turbulence conditions using Gamma-Gamma distribution and negative exponential distribution, respectively \cite{popoola2009bpsk}. However, the authors did not derived closed-form expressions for SIM model. To reveal more insights into SIM, Song et al. studied the error rate performance of SIM using a integration approach and a series expansion of the modified Bessel function of the second kind over the Gamma-Gamma distribution \cite{song2012error}. However, these results are not suitable for USC system, and the error rate performance for USC system using SIM still remains unknown.


In this study, we propose a closed-form NLOS turbulence channel model in terms of Meijer G function over NLOS turbulence channel. Based on this model, we investigate the error rates of SIM model using different modulations in NLOS case. To gain more insights into the USC sytem for SIM model, we use the sum of four hypergeometric functions to represent the NLOS turbulence channel model and use a series form of hypergeometric function. Based on this series form, We obtain closed-form error rate expressions for SIM model. We also analyze the truncation error and derive the asymptotic error rate results. The numerical results demonstrate our Meijer G results and series results are accurate. Furthermore, we study the relationship between the turbulence influence and NLOS transceiver configurations.

This paper is oraganized as follows. In section $\rm{\uppercase\expandafter{\romannumeral2}}$, the SIM model is described. In section $\rm{\uppercase\expandafter{\romannumeral3}}$, we derive an NLOS turbulence channel model. In section $\rm{\uppercase\expandafter{\romannumeral4}}$, we study the error rates of SIM model by using closed-form expressions including Meijer G functions and approximate expressions of finite series terms. Section $\rm{\uppercase\expandafter{\romannumeral5}}$ presents the numerical results and discussions. Finally, Section $\rm{\uppercase\expandafter{\romannumeral6}}$ concludes this work.

\section{SIM Model}
In an optical SIM system, the communication performance depends on the instantaneous signal-to-noise ratio (SNR), which can be written as \cite{song2012error} 
$$
\gamma = C I_s^2.
\eqno{(1)}
$$
The constant $C$ is the average SNR $\overline{\gamma}$ only when $I_s$ is the normalized irradiance. Therefore, if the expectation of the irradiance $E[I_s] \neq 1$, it should be normalized by $I_s/E[I_s]$. After the normalization, we will write the constant $C$ as the average SNR $\overline{\gamma}$ in the following analysis.

\section{NLOS Turbulence Channel Model}
A NLOS communication link consists of two LOS paths assuming single scattering and small common volume \cite{ding2011turbulence}. Specifically, the optical power $p_v$ at the common volume $v$ and the optical power $p_r$ at the receiver can be modeled as \cite{chatzidiamantis2011distribution}
$$
\setlength{\abovedisplayskip}{2pt}
\setlength{\belowdisplayskip}{2pt}
f_{P_v}(p_v) = \frac{2(\alpha_1 \beta_1)^{\frac{\alpha_1+\beta_1}{2}} p_v^{\frac{\alpha_1 + \beta_1}{2}-1}}{\Gamma(\alpha_1) \Gamma(\beta_1) \Omega_v^{\frac{\alpha_1 + \beta_1}{2}}} K_{\alpha_1 - \beta_1} \sqrt{\frac{4\alpha_1 \beta_1}{\Omega_v} p_v },
\eqno{(2)}
$$
$$
\setlength{\abovedisplayskip}{2pt}
\setlength{\belowdisplayskip}{2pt}
f_{P_r|P_v}(p_r|p_v) = \frac{2(\alpha_2 \beta_2)^{\frac{\alpha_2 + \beta_2}{2}} p_r^{\frac{\alpha_2 + \beta_2}{2} -1 }}{ \Gamma(\alpha_2) \Gamma(\beta_2) \Omega_{rv}^{\frac{\alpha_2 + \beta_2}{2}} } K_{\alpha_2 - \beta_2} \sqrt{\frac{4 \alpha_2 \beta_2}{\Omega_{rv}} p_r},
\eqno{(3)}
$$
where the power $p_v > 0$ and $p_r > 0$, $\alpha$ and $\beta$ are respectively the effective numbers of large-scale cells and small-scale cells satisfying $\alpha>\beta$ in OWC systems \cite{wang2010moment}, $\Gamma(\cdot)$ is the gamma function and $K_{\alpha - \beta}(\cdot)$ is the modified Bessel function of the second kind of order $\alpha - \beta$. $\Omega_v$ is the expectation of the power $p_v$ without turbulence, which can be calculated by \cite{shen2019modeling}. $\Omega_{rv}$ is the expectation of the power $p_r$ conditioned on the power $p_v$ without turbulence, which can be obtained from
$$
\setlength{\abovedisplayskip}{2pt}
\setlength{\belowdisplayskip}{2pt}
\Omega_{rv} = {\rm E}(P_r|P_v) = \frac{p_v {\rm exp}(-k_e r_2) A_r}{r_2^{2}}=p_v E_2,
\eqno{(4)}
$$
where $k_e$ is the extinction coefficient and $r_2$ is the distance from the common volume $v$ to the receiver R. Combining the Eq. (2) and Eq. (3), we can derive the probability density function (PDF) of $p_r$. For tractable analysis, we use the transformation of the integration variable along with $t = \sqrt{(4\alpha_1 \beta_1 p_v)/ \Omega_v}$, and the received optical power $p_r$ can be modeled as
$$
\setlength{\abovedisplayskip}{2pt}
\setlength{\belowdisplayskip}{2pt}
f_{P_r}(p_r) =  s p_r^{(\alpha_2 + \beta_2)/2 - 1} \mathcal{I}(p_r),
\eqno{(5)}
$$
where
$$
\setlength{\abovedisplayskip}{2pt}
\setlength{\belowdisplayskip}{2pt}
\mathcal{I}(p_r) = \int_{0}^{\infty} t^{a-1} K_{\alpha_1-\beta_1}(t) \times K_{\alpha_2-\beta_2}(\frac{4 \sqrt{h p_r}}{t}){\rm d }t,
\eqno{(6)}
$$
$$
\setlength{\abovedisplayskip}{2pt}
\setlength{\belowdisplayskip}{2pt}
s = \frac{  2^{ 3-\alpha_1-\beta_1+\alpha_2+\beta_2 } [\alpha_1 \beta_1 \alpha_2 \beta_2 /( \Omega_v E_2 ) ]^{(\alpha_2 + \beta_2)/2}   }{   \Gamma{(\alpha_1)}  \Gamma{(\beta_1)} \Gamma{(\alpha_2)} \Gamma{(\beta_2)} },
\eqno{(7)}
$$
$$
\setlength{\abovedisplayskip}{2pt}
\setlength{\belowdisplayskip}{2pt}
a = \alpha_1+\beta_1-\alpha_2-\beta_2,
\eqno{(8)}
$$
$$
\setlength{\abovedisplayskip}{2pt}
\setlength{\belowdisplayskip}{2pt}
h = { (\alpha_1 \beta_1 \alpha_2 \beta_2) }/{ (\Omega_v E_2) }.
\eqno{(9)}
$$
To derive a closed-form result of the intergation $\mathcal{I}(p_r)$ of Eq. (6), we apply the Mellin convolution theorem \cite{andrews1988integral} to the integration $\mathcal{I}(p_r)$, then the following expression is obtained,
$$
\setlength{\abovedisplayskip}{2pt}
\setlength{\belowdisplayskip}{2pt}
\mathcal{M}[ \mathcal{I}(p_r);s ] = \mathcal{M}[t^a K_{\alpha_1 - \beta_1}(t); s]
\mathcal{M}[K_{\alpha_2 - \beta_2}(t); s].
\eqno{(10)}
$$
Now, applying the results of the Mellin transform of $K_v(ax)$ and $x^af(x)$ \cite{bateman1954tables}, we obtain
$$
\setlength{\abovedisplayskip}{2pt}
\setlength{\belowdisplayskip}{2pt}
\mathcal{M}[t^a K_{\alpha_1 - \beta_1}(t); s] = 2^{s+a-2} \Gamma{(\frac{s+a}{2} - \frac{\alpha_1 - \beta_1}{2} )} 
$$
$$
\setlength{\abovedisplayskip}{2pt}
\setlength{\belowdisplayskip}{2pt}
\times \Gamma{(\frac{s+a}{2} + \frac{\alpha_1 - \beta_1}{2} )} = M_1(s),
\eqno{(11)}
$$
$$
\setlength{\abovedisplayskip}{2pt}
\setlength{\belowdisplayskip}{2pt}
\mathcal{M}[K_{\alpha_2 - \beta_2}(t); s] = 2^{s-2} \Gamma{(  \frac{s}{2}-\frac{\alpha_2-\beta_2}{2} )} 
$$
$$
\setlength{\abovedisplayskip}{2pt}
\setlength{\belowdisplayskip}{2pt}
\times \Gamma{  (   \frac{s}{2}+\frac{\alpha_2-\beta_2}{2}  )  } = M_2(s).
\eqno{(12)}
$$
Substituting the Mellin transform results (11) and (12) into Eq. (10) and applying the inverse Mellin transform \cite{bateman1954tables} to Eq. (10), we can obtain
$$
\setlength{\abovedisplayskip}{2pt}
\setlength{\belowdisplayskip}{2pt}
\mathcal{I}(p_r) = \frac{1}{2 \pi i} \int_{c- i \infty}^{c + i \infty}( 4\sqrt{h p_r} )^{-s} M_1(s) M_2(s){\rm d s}.
\eqno{(13)}
$$
To express the integration $\mathcal{I}(i_r)$ by the Meijer G function form, we transform the integration variable with $s = 2 t$, then the fluctuation of the $p_r$ can be modeled as 
$$
\setlength{\abovedisplayskip}{2pt}
\setlength{\belowdisplayskip}{2pt}
f_{P_r}(p_r) = 2^{a-3} s p_r^{{(\alpha_2+\beta_2)}/{2}-1} 
$$
$$
\setlength{\abovedisplayskip}{2pt}
\setlength{\belowdisplayskip}{2pt}
\times G_{0,4}^{4,0} {\left( {h p_r} \mid{- \atop \frac{2\beta_1 - \alpha_2 - \beta_2}{2}, \frac{ 2\alpha_1 -\alpha_2 -\beta_2 }{2}, \frac{ \beta_2 - \alpha_2 }{2}, \frac{ \alpha_2 - \beta_2 }{2} }   \right)}.
\eqno{(14)}
$$
To reveal more insights, we express Eq. (14) as the sum of four hypergeometric functions, which has been done in \cite{macrobert1953infinite}. Then we use a series expansion for the hypergeometric function \cite{graham1989concrete} and obtain the PDF of the optical power $p_r$ as
$$
\setlength{\abovedisplayskip}{2pt}
\setlength{\belowdisplayskip}{2pt}
g_{P_r}(p_r) = \Xi(\alpha_2 - \beta_2) \sum_{k=0}^{\infty} a_k(\alpha_2-\beta_2) p_{r}^{k+\alpha_2-1}
$$
$$
\setlength{\abovedisplayskip}{2pt}
\setlength{\belowdisplayskip}{2pt}
+\Xi(\beta_2 - \alpha_2) \sum_{k=0}^{\infty} a_k(\beta_2 - \alpha_2) i_{r}^{k+\beta_2-1}
$$
$$
\setlength{\abovedisplayskip}{2pt}
\setlength{\belowdisplayskip}{2pt}
+\Lambda(\alpha_1 - \beta_1) \sum_{k=0}^{\infty} b_k(\alpha_1-\beta_1) p_r^{k+\alpha_1-1} 
$$
$$
\setlength{\abovedisplayskip}{2pt}
\setlength{\belowdisplayskip}{2pt}
+\Lambda(\beta_1-\alpha_1) \sum_{k=0}^{\infty} b_k(\beta_1-\alpha_1) p_r^{k+\beta_1-1},
\eqno{(15)} 
$$
where
$$
\setlength{\abovedisplayskip}{2pt}
\setlength{\belowdisplayskip}{2pt}
\Xi(x) = s 2^{a-2x-3} \Gamma(-x) \Gamma(\frac{ 2 \alpha_1 -\alpha_2 -\beta_2 -x }{2}) 
$$
$$
\setlength{\abovedisplayskip}{2pt}
\setlength{\belowdisplayskip}{2pt}
\times \Gamma(\frac{ 2 \beta_1 -\alpha_2 - \beta_2 - x }{2}) (16h)^{\frac{x}{2}},
\eqno{(16)}
$$
$$
\setlength{\abovedisplayskip}{2pt}
\setlength{\belowdisplayskip}{2pt}
\Theta(x) = \Gamma( 1- \frac{ 2 \alpha_1 -\alpha_2 - \beta_2 - x }{2}) 
$$
$$
\setlength{\abovedisplayskip}{2pt}
\setlength{\belowdisplayskip}{2pt}
\times \Gamma( 1 - \frac{ 2 \beta_1 -\alpha_2 - \beta_2 - x }{2}),
\eqno{(17)}
$$
$$
\setlength{\abovedisplayskip}{2pt}
\setlength{\belowdisplayskip}{2pt}
a_k(x) = \frac{ \Gamma(1+x)  \Theta(x)  h^k   }{ \Gamma(1+x+k) \Theta(x+2k) k!  },
\eqno{(18)}
$$
$$
\setlength{\abovedisplayskip}{2pt}
\setlength{\belowdisplayskip}{2pt}
\Lambda(x) = s 2^{-a - 2x - 3} \Gamma(-x) \Gamma(\frac{  2\alpha_2 - \alpha_1 - \beta_1 - x  }{2})
$$
$$
\setlength{\abovedisplayskip}{2pt}
\setlength{\belowdisplayskip}{2pt}
\times \Gamma( \frac{  2 \beta_2 - \alpha_1 -\beta_1 - x  }{2} ) (16h)^{\frac{a+x}{2}},
\eqno{(19)}
$$
$$
\setlength{\abovedisplayskip}{2pt}
\setlength{\belowdisplayskip}{2pt}
\Psi(x) = \Gamma( 1 + \frac{ \alpha_1 +\beta_1 - 2\alpha_2  + x }{2}) 
$$
$$
\setlength{\abovedisplayskip}{2pt}
\setlength{\belowdisplayskip}{2pt}
\times \Gamma( 1 + \frac{ \alpha_1 +\beta_1 - 2\beta_2  + x}{2}),
\eqno{(20)}
$$
$$
\setlength{\abovedisplayskip}{2pt}
\setlength{\belowdisplayskip}{2pt}
b_k(x) = \frac{  \Gamma(1+x) \Psi(x) h^k }{  \Gamma(1+x+k) \Psi(x+2k) k!  }.
\eqno{(21)}
$$
For tractable analysis, we transform the received optical power $P_r$ in Eq. (14) and (15) to the received irradiance $I_r$ and normalize $I_r$ to be unity by $I_n = (P_r/A_r)/(\Omega_r/A_r)$, where $I_n$ is the normalized irradiance, and 
$$
\setlength{\abovedisplayskip}{2pt}
\setlength{\belowdisplayskip}{2pt}
\Omega_r = E[P_r] = 2^{a-3} s h^{- (\frac{\alpha_2+\beta_2}{2} +1)} \Gamma(\beta_1+1) \Gamma(\beta_2+1) 
$$
$$
\setlength{\abovedisplayskip}{2pt}
\setlength{\belowdisplayskip}{2pt}
\times \Gamma(\alpha_1+1) \Gamma(\alpha_2+1).
\eqno{(22)}
$$
After the normalization, the PDFs of the normalized irradiance transformed from Eq. (14) and (15) are derived as
$$
\setlength{\abovedisplayskip}{2pt}
\setlength{\belowdisplayskip}{2pt}
f_{I_n}(i_n) = 2^{a-3} s \Omega_r^{\frac{\alpha_2+\beta_2}{2}} i_n^{\frac{\alpha_2+\beta_2}{2}-1} 
$$
$$
\setlength{\abovedisplayskip}{2pt}
\setlength{\belowdisplayskip}{2pt}
\times {\rm G}_{0,4}^{4,0} {\left( {h \Omega_r i_n} \mid{- \atop \frac{2\beta_1 - \alpha_2 - \beta_2}{2}, \frac{ 2\alpha_1 -\alpha_2 -\beta_2 }{2}, \frac{ \beta_2 - \alpha_2 }{2}, \frac{ \alpha_2 - \beta_2 }{2} }   \right)},
\eqno{(23)}
$$
and
$$
\setlength{\abovedisplayskip}{2pt}
\setlength{\belowdisplayskip}{2pt}
g_{I_n}(i_n) = \Xi(\alpha_2 - \beta_2) \sum_{k=0}^{\infty} a_k(\alpha_2-\beta_2) \Omega_r^{k+\alpha_2} i_{r}^{k+\alpha_2 - 1}
$$
$$
\setlength{\abovedisplayskip}{2pt}
\setlength{\belowdisplayskip}{2pt}
+\Xi(\beta_2 - \alpha_2) \sum_{k=0}^{\infty} a_k(\beta_2 - \alpha_2) \Omega_r^{k+\beta_2} i_{r}^{k+\beta_2 - 1}
$$
$$
\setlength{\abovedisplayskip}{2pt}
\setlength{\belowdisplayskip}{2pt}
+\Lambda(\alpha_1 - \beta_1) \sum_{k=0}^{\infty} b_k(\alpha_1-\beta_1)  \Omega_r^{k+\alpha_1} i_{r}^{k+\alpha_1-1} 
$$
$$
\setlength{\abovedisplayskip}{2pt}
\setlength{\belowdisplayskip}{2pt}
+\Lambda(\beta_1-\alpha_1) \sum_{k=0}^{\infty} b_k(\beta_1-\alpha_1) \Omega_r^{k+\beta_1} i_{r}^{k+\beta_1-1},
\eqno{(24)}
$$
where the PDF of the normalized irradiance $f_{I_n}(i_n)$ (23) is tranformed from Eq. (14) and $g_{I_n}(i_n)$ (24) is transformed from Eq. (15). We will use the PDFs (23) and (24) of the normalized irradiance $I_n$ in the following performance analysis of SIM for USC system.
\section{Error Rate Analysis Of SIM}
In this section, we study the error rate results for SIM model over a NLOS atmospheric turbulence channel using a direct integration approach, which is given by
$$
\setlength{\abovedisplayskip}{2pt}
\setlength{\belowdisplayskip}{2pt}
P_e = \int_{0}^{\infty}P_e(\overline{\gamma} I_n^2) f_{I_n}(i_n){\rm d} I_n.
\eqno{(25)} 
$$
We will use Eq. (25) to analyze the error rate performance of quadrature phase-shift keying (QPSK) modulation.

\noindent A. QPSK Modulation \par
\noindent (1) Meijer G function representation 

For QPSK modulation, the symbol error rate (SER) of the QPSK modulation can be expressed as $2P(\pi/2)-P(\pi/4)$ \cite{song2012error} where
$$
\setlength{\abovedisplayskip}{2pt}
\setlength{\belowdisplayskip}{2pt}
P(x) = \frac{1}{\pi} \int_{0}^{\infty} \int_{0}^{x}\exp(-\frac{\overline{\gamma_b} I_n^2}{\sin^2\theta}) f_{I_n}(i_n){\rm d}i_n {\rm d}\theta.
\eqno{(26)} 
$$
Substituting the normalized irradiance PDF (23) into Eq. (26) and using the integral transform \cite{functions.wolfram.com}, we obtain
$$
\setlength{\abovedisplayskip}{2pt}
\setlength{\belowdisplayskip}{2pt}
P(x) = A \int_{0}^{x} (\sin\theta)^{\frac{\alpha_2+\beta_2}{2}} G_1(\sin\theta){\rm d}\theta,
\eqno{(27)} 
$$
where 
$$
\setlength{\abovedisplayskip}{2pt}
\setlength{\belowdisplayskip}{2pt}
A = \frac{  2^{2a-7} s \Omega_r^{\frac{ \alpha_2+\beta_2 }{2}}  }{  \pi^3 {\overline{\gamma_b}}^{\frac{ \alpha_2+\beta_2 }{ 4 }}  },
\eqno{(28)} 
$$
\newcounter{mytempeqncnt}
\begin{figure*}[!t]
	\normalsize
	\setcounter{mytempeqncnt}{\value{equation}}
	\setcounter{equation}{5}
	$$
	\setlength{\abovedisplayskip}{2pt}
	\setlength{\belowdisplayskip}{2pt}
	$$
	$$
	\setlength{\abovedisplayskip}{2pt}
	\setlength{\belowdisplayskip}{2pt}
	 {\rm G}_{1,8}^{8,1} {\left( {\frac{(h \Omega_r x)^2}{256\overline{\gamma_b}}} \mid{1-\frac{\alpha_2+\beta_2}{4} \atop \frac{2\beta_1 - \alpha_2 - \beta_2}{4}, \frac{2\beta_1 - \alpha_2 - \beta_2}{4}+\frac{1}{2}, \frac{ 2\alpha_1 -\alpha_2 -\beta_2 }{4}, \frac{ 2\alpha_1 -\alpha_2 -\beta_2 }{4}+\frac{1}{2}, \frac{ \beta_2 - \alpha_2 }{4}, \frac{ \beta_2 - \alpha_2 }{4}+\frac{1}{2}, \frac{ \alpha_2 - \beta_2 }{4} , \frac{ \alpha_2 - \beta_2 }{4} +\frac{1}{2}}   \right)}
	\eqno{(29)} 
	$$
	$$
	\setlength{\abovedisplayskip}{2pt}
	\setlength{\belowdisplayskip}{2pt}
	$$
	$$
	\setlength{\abovedisplayskip}{2pt}
	\setlength{\belowdisplayskip}{2pt}
	{\rm G}_{2,9}^{8,2} {\left( {\frac{(h \Omega_r)^2}{256\overline{\gamma_b}}} \mid{\frac{1}{2}+\frac{\alpha_2+\beta_2}{4}, 1-\frac{\alpha_2+\beta_2}{4} \atop \frac{2\beta_1 - \alpha_2 - \beta_2}{4}, \frac{2\beta_1 - \alpha_2 - \beta_2 + 2}{4}, \frac{ 2\alpha_1 -\alpha_2 -\beta_2 }{4}, \frac{ 2\alpha_1 -\alpha_2 -\beta_2 +2}{4}, \frac{ \beta_2 - \alpha_2 }{4}, \frac{ \beta_2 - \alpha_2 +2}{4}, \frac{ \alpha_2 - \beta_2 }{4} , \frac{ \alpha_2 - \beta_2 +2}{4} , \frac{\alpha_2+\beta_2-4}{4}}   \right)}
	\eqno{(31)} 
	$$
    $$
	\setlength{\abovedisplayskip}{2pt}
	\setlength{\belowdisplayskip}{2pt}
	$$
	$$
	\setlength{\abovedisplayskip}{2pt}
	\setlength{\belowdisplayskip}{2pt}
	{\rm G}_{1,8}^{8,1} {\left( {\frac{(h \Omega_r)^2 j}{256\overline{\gamma_b}}} \mid{1-\frac{\alpha_2+\beta_2}{4} \atop \frac{2\beta_1 - \alpha_2 - \beta_2}{4}, \frac{2\beta_1 - \alpha_2 - \beta_2}{4}+\frac{1}{2}, \frac{ 2\alpha_1 -\alpha_2 -\beta_2 }{4}, \frac{ 2\alpha_1 -\alpha_2 -\beta_2 }{4}+\frac{1}{2}, \frac{ \beta_2 - \alpha_2 }{4}, \frac{ \beta_2 - \alpha_2 }{4}+\frac{1}{2}, \frac{ \alpha_2 - \beta_2 }{4} , \frac{ \alpha_2 - \beta_2 }{4} +\frac{1}{2}}   \right)}
	\eqno{(36)} 
	$$
	\setcounter{equation}{\value{mytempeqncnt}}
	\hrulefill
	\vspace*{4pt}
\end{figure*}
$G_1(x)$ as Eq. (29) is at the top of the next page. Transforming the integration variable $\theta$ into $y$ by $y = \sin^2\theta$ and using the integral formula \cite{functions.wolfram.com}, we obtain the closed-form expression of $P_1(\pi/2)$, which is
$$
\setlength{\abovedisplayskip}{2pt}
\setlength{\belowdisplayskip}{2pt}
P_1(\frac{\pi}{2}) = \frac{A}{2} \Gamma(\frac{1}{2}) G_2,
\eqno{(30)} 
$$
where $G_2$ as Eq. (31) is at the top of the next page. Using Eq. (30) for $P_1(\pi/2)$ and Eq. (27) for $P_1(\pi/4)$, the SER of QPSK modulation  is given by $2P_1(\pi/2) - P_1(\pi/4)$.

\noindent (2) Series representation 

To gain more insights into the SER performance of QPSK modulation, we substitute the PDF of series representation (24) into Eq. (26). Applying Eq. (16) and Eq. (18) in \cite{park2010average} to $P_2({\pi}/{2})$, we obtain
$$
\setlength{\abovedisplayskip}{2pt}
\setlength{\belowdisplayskip}{2pt}
P_2(\frac{\pi}{2}) = \frac{\Xi(\alpha_2-\beta_2)}{4 \pi} \sum_{k=0}^{\infty} a_k(\alpha_2-\beta_2) \Gamma(\frac{k+\alpha_2}{2}) 
$$
$$
\setlength{\abovedisplayskip}{2pt}
\setlength{\belowdisplayskip}{2pt}
\times B(\frac{1}{2}, \frac{k+\alpha_2+1}{2}) \Omega_r^{k+\alpha_2} {\overline{\gamma_b}}^{-\frac{k+\alpha_2}{2}}
$$
$$
\setlength{\abovedisplayskip}{2pt}
\setlength{\belowdisplayskip}{2pt}
 + \frac{\Xi(\beta_2-\alpha_2)}{4 \pi} \sum_{k=0}^{\infty} a_k(\beta_2-\alpha_2)  \Gamma(\frac{k+\beta_2}{2}) 
$$
$$
\setlength{\abovedisplayskip}{2pt}
\setlength{\belowdisplayskip}{2pt}
\times B(\frac{1}{2}, \frac{k+\beta_2+1}{2}) \Omega_r^{k+\beta_2} {\overline{\gamma_b}}^{-\frac{k+\beta_2}{2}} 
$$
$$
\setlength{\abovedisplayskip}{2pt}
\setlength{\belowdisplayskip}{2pt}
+ \frac{\Lambda(\alpha_2-\beta_2)}{4 \pi} \sum_{k=0}^{\infty} b_k(\alpha_2-\beta_2)  \Gamma(\frac{k+\alpha_1}{2}) 
$$
$$
\setlength{\abovedisplayskip}{2pt}
\setlength{\belowdisplayskip}{2pt}
\times B(\frac{1}{2}, \frac{k+\alpha_1+1}{2}) \Omega_r^{k+\alpha_1} {\overline{\gamma_b}}^{-\frac{k+\alpha_1}{2}}
$$
$$
\setlength{\abovedisplayskip}{2pt}
\setlength{\belowdisplayskip}{2pt}
+ \frac{\Lambda(\beta_2-\alpha_2)}{4 \pi} \sum_{k=0}^{\infty} b_k(\beta_2-\alpha_2)  \Gamma(\frac{k+\beta_1}{2}) 
$$
$$
\setlength{\abovedisplayskip}{2pt}
\setlength{\belowdisplayskip}{2pt}
\times B(\frac{1}{2}, \frac{k+\beta_1+1}{2}) \Omega_r^{k+\beta_1} {\overline{\gamma_b}}^{-\frac{k+\beta_1}{2}}.
\eqno{(32)}
$$
Using Eq. (16) and Eq. (18) in \cite{park2010average}, we have
$$
\setlength{\abovedisplayskip}{2pt}
\setlength{\belowdisplayskip}{2pt}
P_2(\frac{\pi}{4}) = \frac{\Xi(\alpha_2-\beta_2)}{2 \pi} \sum_{k=0}^{\infty} a_k(\alpha_2-\beta_2)  \Gamma(\frac{k+\alpha_2}{2}) 
$$
$$
\setlength{\abovedisplayskip}{2pt}
\setlength{\belowdisplayskip}{2pt}
\times g(k+\alpha_2) \Omega_r^{k+\alpha_2} {\overline{\gamma_b}}^{-\frac{k+\alpha_2}{2}}
$$
$$
\setlength{\abovedisplayskip}{2pt}
\setlength{\belowdisplayskip}{2pt}
+ \frac{\Xi(\beta_2-\alpha_2)}{2 \pi} \sum_{k=0}^{\infty} a_k(\beta_2-\alpha_2)  \Gamma(\frac{k+\beta_2}{2}) 
$$
$$
\setlength{\abovedisplayskip}{2pt}
\setlength{\belowdisplayskip}{2pt}
\times g(k+\beta_2) \Omega_r^{k+\beta_2} {\overline{\gamma_b}}^{-\frac{k+\beta_2}{2}} 
$$
$$
\setlength{\abovedisplayskip}{2pt}
\setlength{\belowdisplayskip}{2pt}
+ \frac{\Lambda(\alpha_2-\beta_2)}{2 \pi} \sum_{k=0}^{\infty} b_k(\alpha_2-\beta_2)  \Gamma(\frac{k+\alpha_1}{2}) 
$$
$$
\setlength{\abovedisplayskip}{2pt}
\setlength{\belowdisplayskip}{2pt}
\times g(k+\alpha_1) \Omega_r^{k+\alpha_1} {\overline{\gamma_b}}^{-\frac{k+\alpha_1}{2}}
$$
$$
\setlength{\abovedisplayskip}{2pt}
\setlength{\belowdisplayskip}{2pt}
+ \frac{\Lambda(\beta_2-\alpha_2)}{2 \pi} \sum_{k=0}^{\infty} b_k(\beta_2-\alpha_2)  \Gamma(\frac{k+\beta_1}{2}) 
$$
$$
\setlength{\abovedisplayskip}{2pt}
\setlength{\belowdisplayskip}{2pt}
\times g(k+\beta_1) \Omega_r^{k+\beta_1} {\overline{\gamma_b}}^{-\frac{k+\beta_1}{2}},
\eqno{(33)}
$$
where $g(x)$ is defined in \cite{song2012error}. With Eq. (32) and Eq. (33), the SER of QPSK modulation can be given by $2P_2(\pi/2)-P_2(\pi/4)$. In the following, we will derive the error rate of diffenential phase-shift keying and noncoherent frequency-shift keying (NCFSK) modulation in NLOS case.

\noindent B. DPSK/NCFSK Modulation\par
\noindent (1) Meijer G function representation 

For DPSK and NCFSK modulation, the conditional bit error rate (BER) is $P_{e,j}(\gamma_b) = (1/2)\exp[-\gamma_b/(j)]$ \cite{sklar1988digital} where $j=1$ for DPSK and $j=2$ for NCFSK. Using Eq. (25), we can obtain
$$
\setlength{\abovedisplayskip}{2pt}
\setlength{\belowdisplayskip}{2pt}
P_{e,j}=\frac{1}{2} \int_{0}^{\infty} \exp(-\frac{\overline{\gamma_b}}{j} I_n^2)f_{I_n}(i_n){\rm d}i_n.
\eqno{(34)}
$$
Substituting Eq. (23) into Eq. (34) and using the integral transform \cite{functions.wolfram.com}, we have
$$
\setlength{\abovedisplayskip}{2pt}
\setlength{\belowdisplayskip}{2pt}
P_{e,j} = \frac{4^{a-4} s \Omega_r^{\frac{\alpha_2+\beta_2}{2}}}{\pi^2} (\frac{\overline{\gamma_b}}{j})^{-\frac{\alpha_2+\beta_2}{4}} G_3,
\eqno{(35)}
$$
where $G_3$ as Eq. (36) is at the top of the this page.

\noindent (2) Series representation 

Using Eq. (16) and Eq. (18) \cite{park2010average} and substituting Eq. (24) into Eq. (34), we obtain
$$
\setlength{\abovedisplayskip}{2pt}
\setlength{\belowdisplayskip}{2pt}
P_{e,j} = \frac{\Xi(\alpha_2-\beta_2)}{4} \sum_{k=0}^{\infty} a_k(\alpha_2-\beta_2)  
$$
$$
\setlength{\abovedisplayskip}{2pt}
\setlength{\belowdisplayskip}{2pt}
\times \Gamma(\frac{k+\alpha_2}{2})  \Omega_r^{k+\alpha_2} (\frac{\overline{\gamma_b}}{j})^{-\frac{k+\alpha_2}{2}}
$$
$$
\setlength{\abovedisplayskip}{2pt}
\setlength{\belowdisplayskip}{2pt}
+ \frac{\Xi(\beta_2-\alpha_2)}{4} \sum_{k=0}^{\infty} a_k(\beta_2-\alpha_2)   
$$
$$
\setlength{\abovedisplayskip}{2pt}
\setlength{\belowdisplayskip}{2pt}
\times \Gamma(\frac{k+\beta_2}{2}) \Omega_r^{k+\beta_2} (\frac{\overline{\gamma_b}}{j})^{-\frac{k+\beta_2}{2}} 
$$
$$
\setlength{\abovedisplayskip}{2pt}
\setlength{\belowdisplayskip}{2pt}
+ \frac{\Lambda(\alpha_2-\beta_2)}{4} \sum_{k=0}^{\infty} b_k(\alpha_2-\beta_2)  
$$
$$
\setlength{\abovedisplayskip}{2pt}
\setlength{\belowdisplayskip}{2pt}
\times \Gamma(\frac{k+\alpha_1}{2})  \Omega_r^{k+\alpha_1} (\frac{\overline{\gamma_b}}{j})^{-\frac{k+\alpha_1}{2}}
$$
$$
\setlength{\abovedisplayskip}{2pt}
\setlength{\belowdisplayskip}{2pt}
+ \frac{\Lambda(\beta_2-\alpha_2)}{4} \sum_{k=0}^{\infty} b_k(\beta_2-\alpha_2)   
$$
$$
\setlength{\abovedisplayskip}{2pt}
\setlength{\belowdisplayskip}{2pt}
\times \Gamma(\frac{k+\beta_1}{2}) \Omega_r^{k+\beta_1} (\frac{\overline{\gamma_b}}{j})^{-\frac{k+\beta_1}{2}}.
\eqno{(37)}
$$
In this part, we derive the error rate results of QPSK modulation and DPSK/NCFSK modulation both in Meijer G function form and series form. 

\noindent C. Truncation Error Analysis

Since we only use finite terms of the series results, we have to analyze the truncation error first. By using the first $J+1$ terms in Eq. (32), we derive the trunction error for $P_2(\pi/2)$ as 
$$
\setlength{\abovedisplayskip}{2pt}
\setlength{\belowdisplayskip}{2pt}
\epsilon_1(J) = \frac{1 }{4 \pi} \sum_{k=J+1}^{\infty} \frac{1}{k!} (\frac{h \Omega_r}{\sqrt{\overline{\gamma_b}}})^k  
$$
$$
\setlength{\abovedisplayskip}{2pt}
\setlength{\belowdisplayskip}{2pt}
\times [ \Xi(\alpha_2-\beta_2) v_k(\alpha_2,\alpha_2-\beta_2,\alpha_2-\alpha_1,\alpha_2-\beta_1)
$$
$$
\setlength{\abovedisplayskip}{2pt}
\setlength{\belowdisplayskip}{2pt}
+\Xi(\beta_2-\alpha_2) v_k(\beta_2, \beta_2-\alpha_2, \beta_2-\alpha_1, \beta_2-\beta_1)
$$
$$
\setlength{\abovedisplayskip}{2pt}
\setlength{\belowdisplayskip}{2pt}
+\Lambda(\alpha_1-\beta_1)v_k(\alpha_1, \alpha_1-\beta_1, \alpha_1-\alpha_2, \alpha_1-\beta_2)
$$
$$
\setlength{\abovedisplayskip}{2pt}
\setlength{\belowdisplayskip}{2pt}
+\Lambda(\beta_1-\alpha_1)v_k(\beta_1, \beta_1-\alpha_1, \beta_1-\alpha_2, \beta_1-\beta_2) ],
\eqno{(38)}
$$
where
$$
\setlength{\abovedisplayskip}{2pt}
\setlength{\belowdisplayskip}{2pt}
v_k(w,x,y,z) =I_k(w,x,y,z)L(w,x,y,z){\rm B}(\frac{1}{2}, \frac{k+w+1}{2}),
\eqno{(39)}
$$
$$
\setlength{\abovedisplayskip}{2pt}
\setlength{\belowdisplayskip}{2pt}
I_k(w,x,y,z) = \frac{\Gamma[(k+w)/2]}{\Gamma(1+x+k)\Gamma(1+y+k)\Gamma(1+z+k)},
\eqno{(40)}
$$
$$
\setlength{\abovedisplayskip}{2pt}
\setlength{\belowdisplayskip}{2pt}
L(w,x,y,z) = \Gamma(1+x)\Gamma(1+y)\Gamma(1+z) (\frac{\Omega_r}{\sqrt{\overline{\gamma_b}}})^w.
\eqno{(41)}
$$
We can obtain an upper bound of the truncation error by substituting the summation into the exponential function
$$
\setlength{\abovedisplayskip}{2pt}
\setlength{\belowdisplayskip}{2pt}
\epsilon_1(J) \leq \frac{1 }{4 \pi} \exp(\frac{h \Omega_r }{\overline{\sqrt{\gamma_b}}}) \max_{k>J}V_k(\alpha_1,\beta_1,\alpha_2,\beta_2),
\eqno{(42)} 
$$
where 
$$
\setlength{\abovedisplayskip}{2pt}
\setlength{\belowdisplayskip}{2pt}
V_k(\alpha_1,\beta_1,\alpha_2,\beta_2) = \Xi(\alpha_2-\beta_2) v_k(\alpha_2,\alpha_2-\beta_2,\alpha_2-\alpha_1,\alpha_2-\beta_1)
$$
$$
\setlength{\abovedisplayskip}{2pt}
\setlength{\belowdisplayskip}{2pt}
+\Xi(\beta_2-\alpha_2) v_k(\beta_2, \beta_2-\alpha_2, \beta_2-\alpha_1, \beta_2-\beta_1)
$$
$$
\setlength{\abovedisplayskip}{2pt}
\setlength{\belowdisplayskip}{2pt}
+\Lambda(\alpha_1-\beta_1)v_k(\alpha_1, \alpha_1-\beta_1, \alpha_1-\alpha_2, \alpha_1-\beta_2)
$$
$$
\setlength{\abovedisplayskip}{2pt}
\setlength{\belowdisplayskip}{2pt}
+\Lambda(\beta_1-\alpha_1)v_k(\beta_1, \beta_1-\alpha_1, \beta_1-\alpha_2, \beta_1-\beta_2). 
\eqno{(43)}
$$
Fixing the value of $w, x, y, z$, we observe that the second term $L(w,x,y,z)$ and the third term ${\rm B}({1}/{2}, {(k+w+1)}/{(2)})$ are finite and the first term $I_k(w,x,y,z)$ approaches zero when index $k$ approaches $\infty$. According to the infinitesimal properties, we observe that $V_k(\alpha_1, \beta_1, \alpha_2, \beta_2)$ approaches zero when index $k$ approach $\infty$. We note that $V_k(\alpha_1,\beta_1,\alpha_2,\beta_2)$ is a number sequence of discrete numbers, which limits to zero. Therefore, we conclude that the maximum value of $V_k(\alpha_1,\beta_1,\alpha_2,\beta_2)$ exists.

After investigating Eq. (33), we derive an lower bound of the truncation error for $P(\pi/4)$ as
$$
\setlength{\abovedisplayskip}{2pt}
\setlength{\belowdisplayskip}{2pt}
\epsilon_2(J) \geq \frac{1 }{2 \pi} \exp(\frac{h \Omega_r }{\overline{\sqrt{\gamma_b}}}) \min_{k>J} U_k(\alpha_1,\beta_1,\alpha_2,\beta_2),
\eqno{(44)}
$$
where
$$
\setlength{\abovedisplayskip}{2pt}
\setlength{\belowdisplayskip}{2pt}
U_k(\alpha_1,\beta_1,\alpha_2,\beta_2) = \Xi(\alpha_2-\beta_2) u_k(\alpha_2,\alpha_2-\beta_2,\alpha_2-\alpha_1,\alpha_2-\beta_1)
$$
$$
\setlength{\abovedisplayskip}{2pt}
\setlength{\belowdisplayskip}{2pt}
+\Xi(\beta_2-\alpha_2) u_k(\beta_2, \beta_2-\alpha_2, \beta_2-\alpha_1, \beta_2-\beta_1)
$$
$$
\setlength{\abovedisplayskip}{2pt}
\setlength{\belowdisplayskip}{2pt}
+\Lambda(\alpha_1-\beta_1)u_k(\alpha_1, \alpha_1-\beta_1, \alpha_1-\alpha_2, \alpha_1-\beta_2)
$$
$$
\setlength{\abovedisplayskip}{2pt}
\setlength{\belowdisplayskip}{2pt}
+\Lambda(\beta_1-\alpha_1)u_k(\beta_1, \beta_1-\alpha_1, \beta_1-\alpha_2, \beta_1-\beta_2),
\eqno{(45)}
$$
$$
\setlength{\abovedisplayskip}{2pt}
\setlength{\belowdisplayskip}{2pt}
u_k(w,x,y,z) = I_k(w,x,y,z) L(w,x,y,z) g(k+w).
\eqno{(46)}
$$
Similar to $V_k(\alpha_1, \beta_1, \alpha_2, \beta_2)$, the minmum value of $U_k(\alpha_1, \beta_1, \alpha_2, \beta_2)$ exists. Using Eq. (42) and Eq. (44), we obtain the truncation error of $2P(\pi/2)-P(\pi/4)$ as $2\epsilon_1(J)-\epsilon_2(J)$, which is found to approach zero when index $k$ approaches $\infty$ based on infinitesimal properties. In addition, the truncation error of Eq. (37) can also be found following the derivation of Eq. (42) and Eq. (44). 
\setcounter{figure}{-1} 
\begin{figure*}[h]
	\centering
	
	\subfloat{
		\begin{minipage}[c][0.95\totalheight][b]{1\columnwidth}%
			\setlength{\abovecaptionskip}{0.cm}
			\setlength{\belowcaptionskip}{-0.cm}
			\centering
			\includegraphics[width=3.1713in,height=2.3213in]{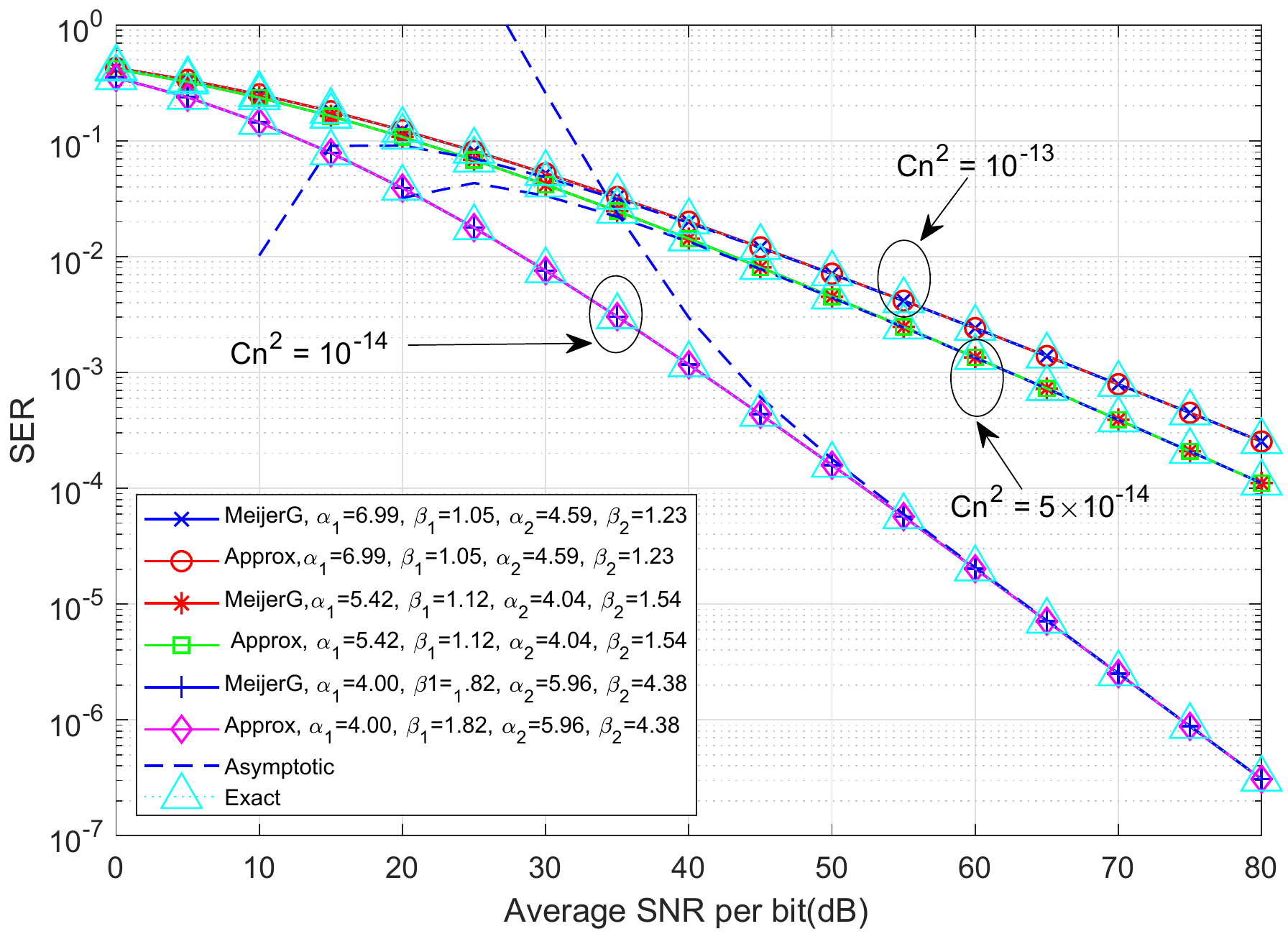}
			\caption{SERs of QPSK modulated USC system over NLOS turbulence channels with $J=30$. The Meijer G results are obtained from Eqs. (27) and (30), while the approximate results are obtained from Eqs. (32) and (33).}
			\label{Fig.1}
		\end{minipage}
	}%
	\subfloat{
		\begin{minipage}[c][0.95\totalheight][b]{1\columnwidth}%
			\setlength{\abovecaptionskip}{0.cm}
			\setlength{\belowcaptionskip}{-0.cm}
			\centering
			\includegraphics[width=3.1713in,height=2.3213in]{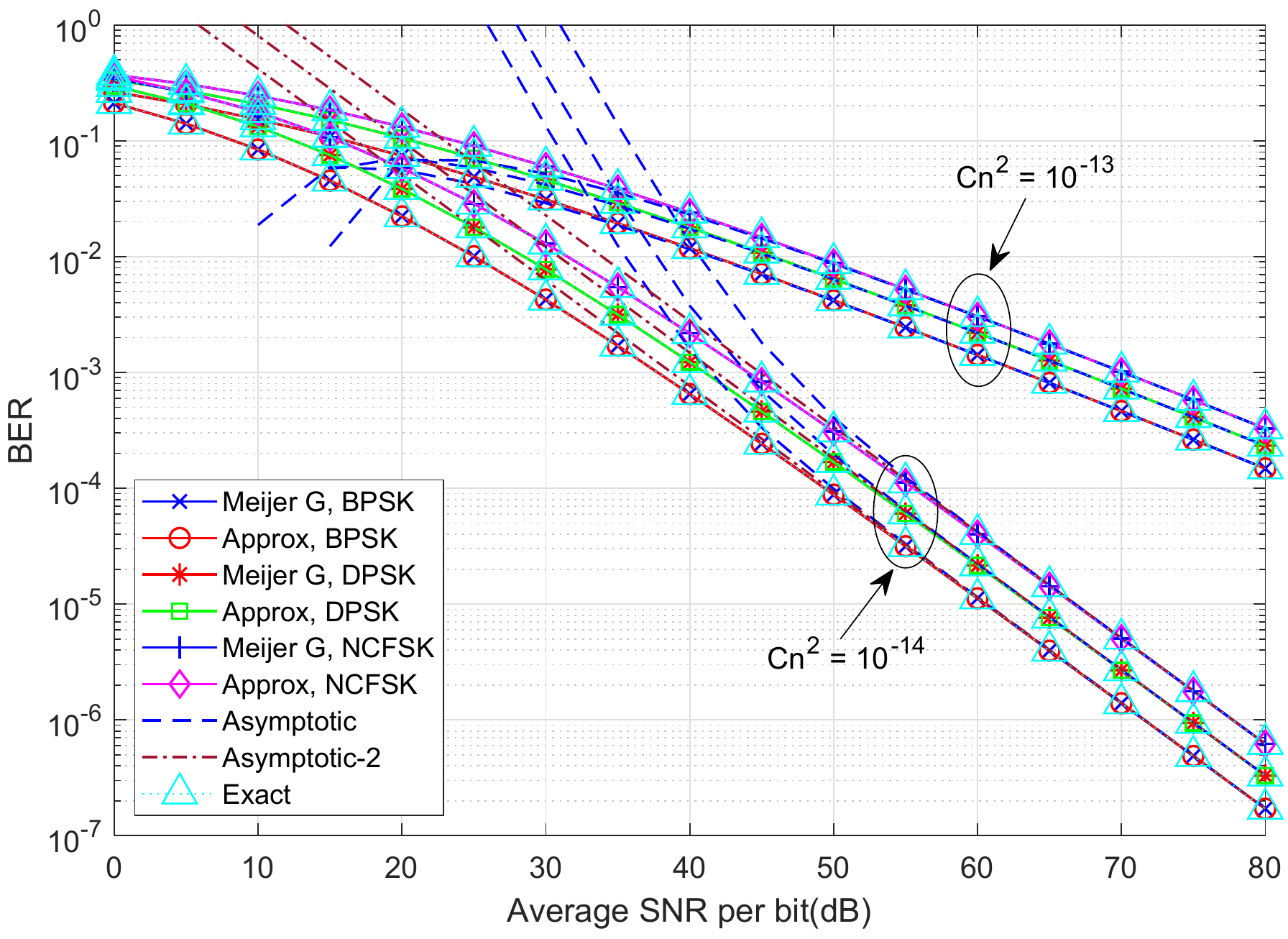}
			\caption{BERs of USC system using BPSK, DPSK and NCFSK over NLOS turbulence channels with $\alpha_1=6.99, \beta_1=1.05, \alpha_2=4.59, \beta_2=1.23$ and $J=30$. The Meijer G results are obtained from Eqs. (30) and (35), while the approximate results are obtained from Eqs. (32) and (37).}
			\label{Fig.2}
		\end{minipage}
	}%
	
	\subfloat{
		\begin{minipage}[c][0.95\totalheight][b]{1\columnwidth}%
			\setlength{\abovecaptionskip}{0.cm}
			\setlength{\belowcaptionskip}{-0.cm}
			\centering
			\includegraphics[width=3.1713in,height=2.3213in]{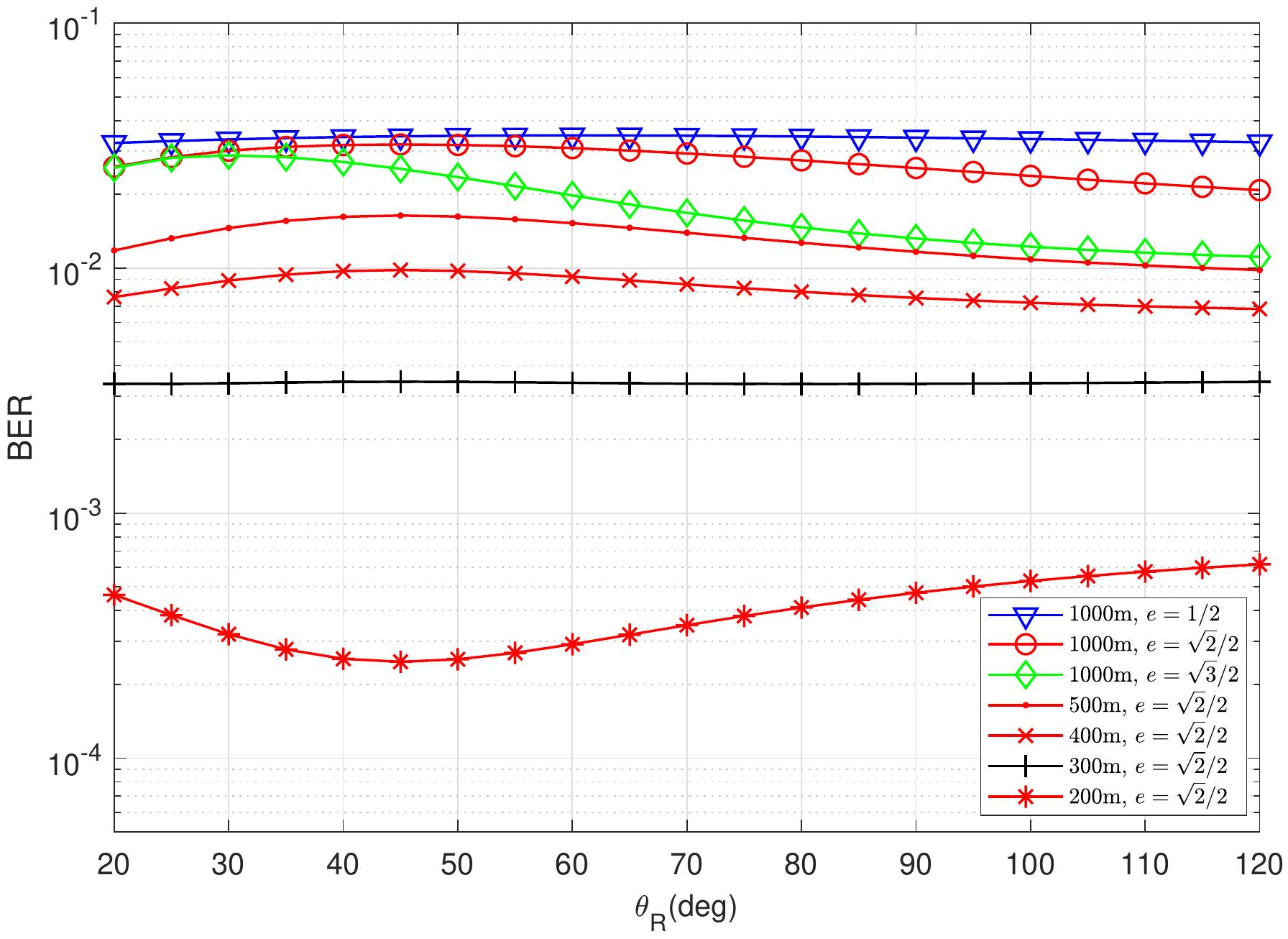}
			\caption{BERs of BPSK modulated USC system for different ellipse settings at different communication ranges under strong turbulence condition of ${\rm Cn}^2=10^{-13} {\rm m}^{-2/3}$.}
	        \label{Fig.3}
		\end{minipage}
	}%
	\subfloat{
		\begin{minipage}[c][0.95\totalheight][b]{1\columnwidth}%
			\setlength{\abovecaptionskip}{0.cm}
			\setlength{\belowcaptionskip}{-0.cm}
			\centering
			\includegraphics[width=3.1713in,height=2.3213in]{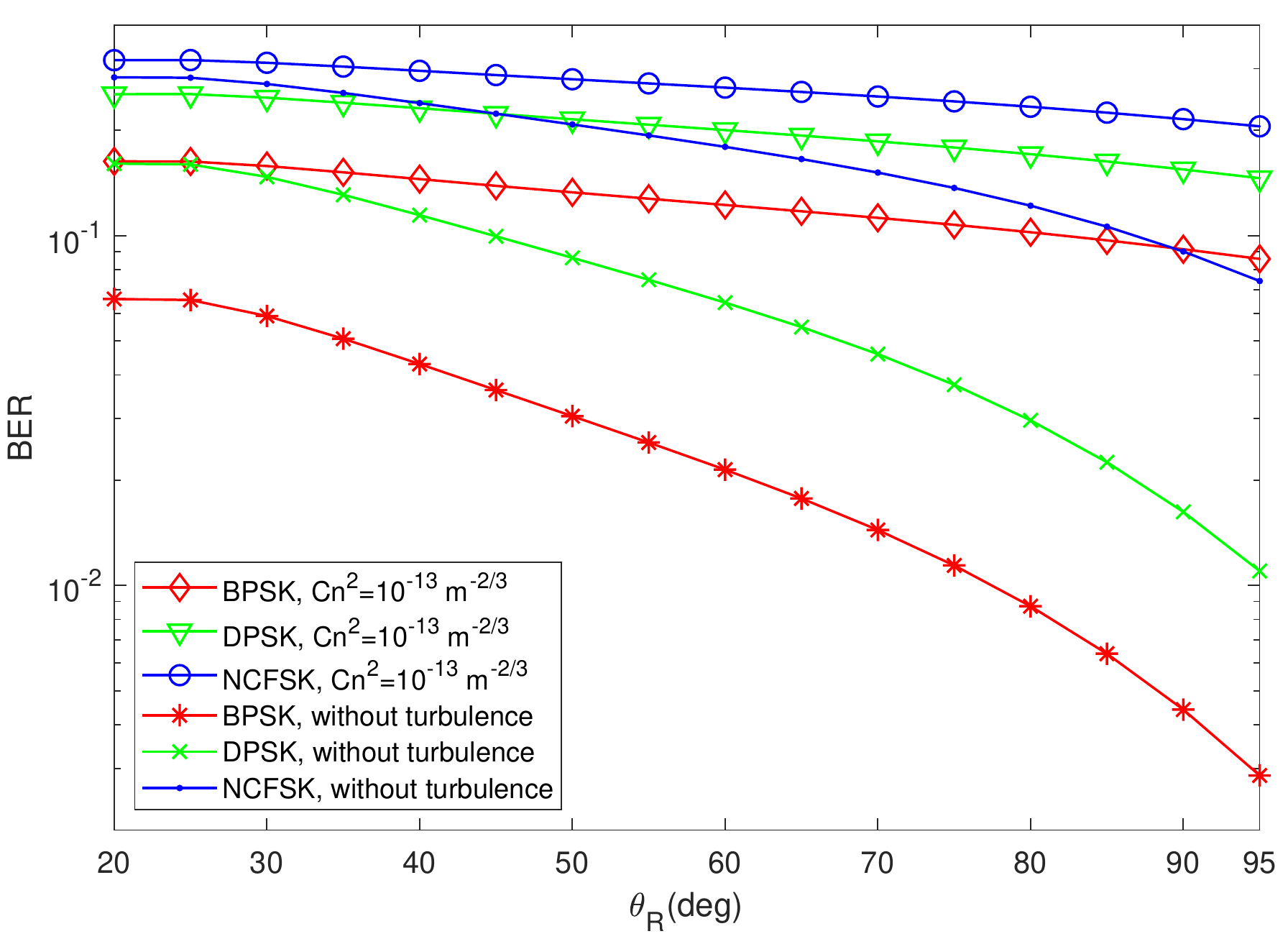}
			\caption{BERs of BPSK, DPSK, and NCFSK modulated USC system for an ellipse of $e=\sqrt{2}/2$ with and without turbulence at a range of $200\,\rm{m}$.}
			\label{Fig.4}
		\end{minipage}
	}%
	
	\centering
\end{figure*}

\noindent D. Asymptotic Error Analysis

We note that the relationship $\alpha>\beta$ always holds in optical communication scenarios, which indicates that the term $(\overline{\gamma_b})^{-(k+\alpha_2)/2}$ diminishes faster than the term $(\overline{\gamma_b})^{-(k+\beta_2)/2}$ and the term $(\overline{\gamma_b})^{-(k+\alpha_1)/2}$ diminishes faster than the term $(\overline{\gamma_b})^{-(k+\beta_1)/2}$. Therefore, the leading terms in Eq. (32) become the dominant terms in high $\overline{\gamma_b}$, which suggests the error rate for BPSK in NLOS case is
$$
\setlength{\abovedisplayskip}{2pt}
\setlength{\belowdisplayskip}{2pt}
P^{'}_{b} = \frac{s 2^{\alpha_1+\beta_1+\alpha_2-3\beta_2-3} \Gamma(\alpha_2-\beta_2) \Gamma(\alpha_1-\beta_2) \Gamma(\beta_1-\beta_2)}{4 \pi}
$$
$$
\setlength{\abovedisplayskip}{2pt}
\setlength{\belowdisplayskip}{2pt}
\times (16h)^{-\frac{\alpha_2-\beta_2}{2}} \Gamma(\frac{\beta_2}{2}) {\rm B}(\frac{1}{2},\frac{\beta_2+1}{2})  \Omega_r^{\beta_2} \overline{\gamma_b}^{-\frac{\beta_2}{2}} 
$$
$$
\setlength{\abovedisplayskip}{2pt}
\setlength{\belowdisplayskip}{2pt}
+ \frac{s 2^{\alpha_1-3 \beta_1+\alpha_2+\beta_2-3} \Gamma(\alpha_1-\beta_1) \Gamma(\alpha_2-\beta_1) \Gamma(\beta_2-\beta_1)}{4 \pi}
$$
$$
\setlength{\abovedisplayskip}{2pt}
\setlength{\belowdisplayskip}{2pt}
\times (16h)^{\frac{2 \beta_1 -\alpha_2 - \beta_2}{2}} \Gamma(\frac{\beta_1}{2}) {\rm B}(\frac{1}{2},\frac{\beta_1+1}{2})  \Omega_r^{\beta_1} \overline{\gamma_b}^{-\frac{\beta_1}{2}}
$$
$$
\setlength{\abovedisplayskip}{2pt}
\setlength{\belowdisplayskip}{2pt}
=P^{'}_b(\beta_1)+P^{'}_b(\beta_2).
\eqno{(47)}
$$
Similarly, the error rate for DPSK/NCFSK is 
$$
\setlength{\abovedisplayskip}{2pt}
\setlength{\belowdisplayskip}{2pt}
P^{'}_{e,k} = \frac{s 2^{\alpha_1+\beta_1+\alpha_2-3\beta_2-3} \Gamma(\alpha_2-\beta_2) \Gamma(\alpha_1-\beta_2) \Gamma(\beta_1-\beta_2)}{4}
$$
$$
\setlength{\abovedisplayskip}{2pt}
\setlength{\belowdisplayskip}{2pt}
\times (16h)^{-\frac{\alpha_2-\beta_2}{2}} \Gamma(\frac{\beta_2}{2}) \Omega_r^{\beta_2} (\frac{\overline{\gamma_b}}{j})^{-\frac{\beta_2}{2}} 
$$
$$
\setlength{\abovedisplayskip}{2pt}
\setlength{\belowdisplayskip}{2pt}
+ \frac{s 2^{\alpha_1-3\beta_1+\alpha_2+\beta_2-3} \Gamma(\alpha_1-\beta_1) \Gamma(\alpha_2-\beta_1) \Gamma(\beta_2-\beta_1)}{4}
$$
$$
\setlength{\abovedisplayskip}{2pt}
\setlength{\belowdisplayskip}{2pt}
\times (16h)^{\frac{2\beta_1-\alpha_2-\beta_2}{2}} \Gamma(\frac{\beta_1}{2})\Omega_r^{\beta_1} (\frac{\overline{\gamma_b}}{j})^{-\frac{\beta_1}{2}}.
$$
$$
\setlength{\abovedisplayskip}{2pt}
\setlength{\belowdisplayskip}{2pt}
=P^{'}_{e,j}(\beta_1)+P^{'}_{e,j}(\beta_2).
\eqno{(48)}
$$
 From Eq. (47) and Eq. (48), we can find the SNR penalty factor by bisection method for a given error rate. Furthermore, Eq. (47) and Eq. (48) can be simplified by $P^{'}_b(\beta)$ where $\beta={\rm min}{(\beta_1,\beta_2)}$. In this case, the SNR penalty factor between BPSK and DPSK/NCFSK are
$$
\setlength{\abovedisplayskip}{2pt}
\setlength{\belowdisplayskip}{2pt}
{\rm SNR}_{{\rm BPSK}-{\rm DPSK}} = \frac{20}{\beta}{\rm log}[\frac{\pi j^{\frac{\beta}{2}}}{ {\rm B}(\frac{1}{2},\frac{\beta+1}{2})}],
\eqno{(49)}
$$
$$
\setlength{\abovedisplayskip}{2pt}
\setlength{\belowdisplayskip}{2pt}
{\rm SNR}_{{\rm DPSK}-{\rm NCFSK}} = 10{\rm log}2,
\eqno{(50)}
$$
where $\log(\cdot)$ is the logarithm with the base 10. We note that ${\rm SNR}_{{\rm BPSK}-{\rm DPSK}}$ only depends on the smaller channel parameter $\beta$ and ${\rm SNR}_{{\rm DPSK}-{\rm NCFSK}}$ is constant.

\section{Numerical results}
In this section, we compare the Meijer G form error rate and the approximate error rate of finite terms with the exact error rate to validate the derived results. The Meijer G form error rate results are derived by using integrals including Meijer G functions. The approximate error rate results are obtained by using first $J+1$ terms, and the exact error rate results are evaluated by numerical integration. We consider three turbulence conditions with ${\rm Cn}^2=10^{-14} {\rm m}^{-2/3}$ and ${\rm Cn}^2=5\times10^{-14} {\rm m}^{-2/3}$ as moderate and ${\rm Cn}^2=10^{-13} {\rm m}^{-2/3}$ as strong turbulence. We also evaluate the error rate results for different NLOS transceiver configurationss, especially for transceiver elevation angles. The simulation parameters are shown in Tab. 1.
\begin{table}[]
	\caption{Simulation parameters}
	\vspace{1pt}
	\centering
	\begin{tabular}{p{2cm}p{3cm}p{2.5cm}}
		\hline
		Parameters & Value   \\
		\hline                                                    
		$k_a$     & $0.802 ~{\rm km}^{-1}$                       \\
		$k_r$     & $0.266 ~{\rm km}^{-1}$                       \\
		$k_m$     & $0.284 ~{\rm km}^{-1}$                       \\
		$\gamma$  & $0.017$                                      \\
		$g$       & $0.72$                                       \\
		$f$       & $0.5$                                        \\
	    $A_r$     & $1.77\times 10^{-4}~{\rm m}^2$               \\  
		$\lambda$ &$260~{\rm nm}$                                \\
		\hline       
	\end{tabular}
	\label{bs2}
\end{table}

In Fig. 1, we present SERs for QPSK modulated USC system for different ${\rm Cn}^2$ values. The transceiver configurations are set as: transmitter elevation angle $\theta_T=30^{\circ}$, transmitter beam angle $\beta_T = 8~{\rm mrad}$, receiver elevation angle $\theta_R = 80^{\circ}$, reveiver filed of view angle $\beta_R=20^{\circ}$ and baseline distance $r = 1000~{\rm m}$ between the transmitter and the receiver. The Meijer G form results are obtained by Eqs. (27) and (30), while the approximate results are obtained by Eqs. (32) and (33). The presented results agree well with each other. From Fig. 1, we note that the asymptotic SERs agree well with the exact SERs for different ${\rm Cn^2}$ values in high SNR values. In particular, we comment that when $\beta_1$ and $\beta_2$ are smaller, which corresponds to ${\rm Cn^2}=10^{-13} {\rm m}^{-2/3}$ for stronger turbulence condition, the asymptotic SERs will approach faster to the exact SERs. Furthermore, we note that Fig. 1 does not show the asymptotic results in low SNR regimes since the asymptotic results are bigger than one for ${\rm Cn^2}=10^{-14} {\rm m}^{-2/3}$, and the asymptotic results are negative for ${\rm Cn^2}=10^{-13} {\rm m}^{-2/3}$. 

In Fig. 2, BERs of USC system using SIM are presented over NLOS turbulence channel with different ${\rm Cn^2}$ values using BPSK, DPSK, and NCFSK modulations. The transceiver configurations are set as: $(\theta_T, \beta_T, \theta_R, \beta_R, r) = (30^{\circ}, 8~{\rm mrad}, 80^{\circ}, 20^{\circ}, 1000~{\rm m})$. The Meijer G form results are obtained by Eqs. (30) and (35), while the approximate results are obtained by Eqs. (32) and (37). From Fig. 2, we again observe that the derived Meijer G form results and the approximate results with $J=30$ conform to the exact results. For strong turbulence condition of ${\rm Cn^2}=10^{-13} {\rm m}^{-2/3}$, we note that the terms $\overline{\gamma_b}^{-(1.05)/(2)}$ and $\overline{\gamma_b}^{-(1.23)/(2)}$ are domiant, therefore, we use Eqs. (47) and (48) to obtain the asymptotic results. In particular, when the BER level is at $10^{-3}$, we use bisection method to solve Eqs. (47) and (48) and find that ${\rm SNR}_{{\rm BPSK}-{\rm DPSK}}$ is $3.98~{\rm dB}$ and ${\rm SNR}_{{\rm DPSK}-{\rm NCFSK}}$ is $3.01~{\rm dB}$, which agree with the SNR penalty factor results of $3.98~{\rm dB}$ and $3.05~{\rm dB}$ respectively from Fig. 2. For moderate turbulence condition of ${\rm Cn^2}=10^{-14} {\rm m}^{-2/3}$, we observe that the term $\overline{\gamma_b}^{-(1.82)/(2)}$ becomes domiant. Therefore, we use $P^{'}_b(\beta_1)$ in Eq. (47) and $P^{'}_{e,j}(\beta_1)$ in Eq. (48) to obtain the asymptotic-2 results. From Fig. 2, we can see that the asymptotic-2 results agree with the exact results in high SNR regimes, which indicates using a single parameter $\beta$ is sufficient to estimate error rates in this case. Then we use Eqs. (50) and (51) to obtain the SNR penalty factors. Particularly, when the BER level is at $10^{-6}$, we find from Fig. 2 that ${\rm SNR}_{{\rm BPSK}-{\rm DPSK}}$ is $3.13~{\rm dB}$ and ${\rm SNR}_{{\rm DPSK}-{\rm NCFSK}}$ is $3.01~{\rm dB}$, which agree with the calculation results of $3.19~{\rm dB}$ and $2.96~{\rm dB}$ respectively from Eqs. (49) and (50).

The previous studies have shown that ultraviolet scattering communications have different path loss for different NLOS tranceiver configurations \cite{shen2019modeling}, which motivates us to study the influence of turbulence to different NLOS tranceiver configurations, especially for different tranceiver elevation angles. Based on this motivation, we assume the same SNR of $30~{\rm dB}$ and keep the same NLOS transmit distances, which can be achieved by setting the transmitter and the receiver on the focal points of a given ellipse and setting the intersection point of the transmitter axis and the receiver axis on the ellipse. 

In Fig. 3, we present BERs of USC system using SIM for different ellipse settings at different communication ranges. The transceiver configurations are set as: $(\beta_T, \beta_R) = (8~{\rm mrad}, 20^{\circ})$. From Fig. 3, we observe that at a the communication range of 1000m, when the eccentricities $e$ decrease, the BERs increase and change slowly due to larger propagation distance. We also note that the error rate results have a maximum value at $\theta_R = 30^{\circ}$ for the communication range of 1000m and the eccentricity of $\sqrt{3}/2$, which corresponds to the case of the same distance of two LOS paths. The similar results can be observed for other eccentricity settings. Furthermore, we comment that when the communication range decrease from 1000m to 300m, the BERs decrease and change slowly with a maximum value for eccentricity of $\sqrt{2}/2$. However, when the communication range decrease to 200m, the BERs contrarily have a minimum value when two-LOS link formulates the same distance, since the parameter $\alpha$ is higher for each LOS path. 

We have assumed the same average SNR in Fig. 3, however, average SNR is a function of the received power, which can can derived as \cite{agrawal2012fiber}
$$
\overline{\gamma_b} = \frac{\eta_f \eta_r \lambda P_r}{h c B}
\eqno{(51)}
$$
for photomultiplier, where $\eta_f$ is the filter transmission, $\eta_r$ is the detector quantum efficiency, $\lambda$ is the wavelength, $h$ is plank constant, $c$ is the speed of light, $B$ is the bit rate, and $P_r$ is the received power caclulated by \cite{shen2019modeling}. We set $\eta_f = 0.1$ \cite{xu2008analytical}, $\eta_r = 0.2$ \cite{xu2008analytical} and $B = 5000~{\rm bit/s}$. In Fig. 4, BERs of BPSK, DPSK, and NCFSK modulated UV communication system are presented with and without turbulence. The transceiver configurations are set as: $(\beta_T, \beta_R, r, e) = (8~{\rm mrad}, 20^{\circ}, 200~{\rm m}, \frac{\sqrt{2}}{2})$.

In Fig. 4, comparing the results of USC system with and without turbulence, we observe that the error rate performance is seriously damaged by turbulence of $\rm{Cn^2}=10^{-13} {\rm m}^{-2/3}$. For example, when $\theta_R$ varies from $20^{\circ}$ to $95^{\circ}$, the BERs varies form $0.0659$ to $2.87\times10^{-3}$ without turbulence and $0.163$ to $8.58\times10^{-2}$ with turbulence.
\setcounter{figure}{4} 
\begin{figure}[htbp]
	\begin{minipage}[c][0.95\totalheight][b]{1\columnwidth}%
		\setlength{\abovecaptionskip}{0.cm}
		\setlength{\belowcaptionskip}{-0.cm}
		\centering
		\includegraphics[width=3.1713in,height=2.3213in]{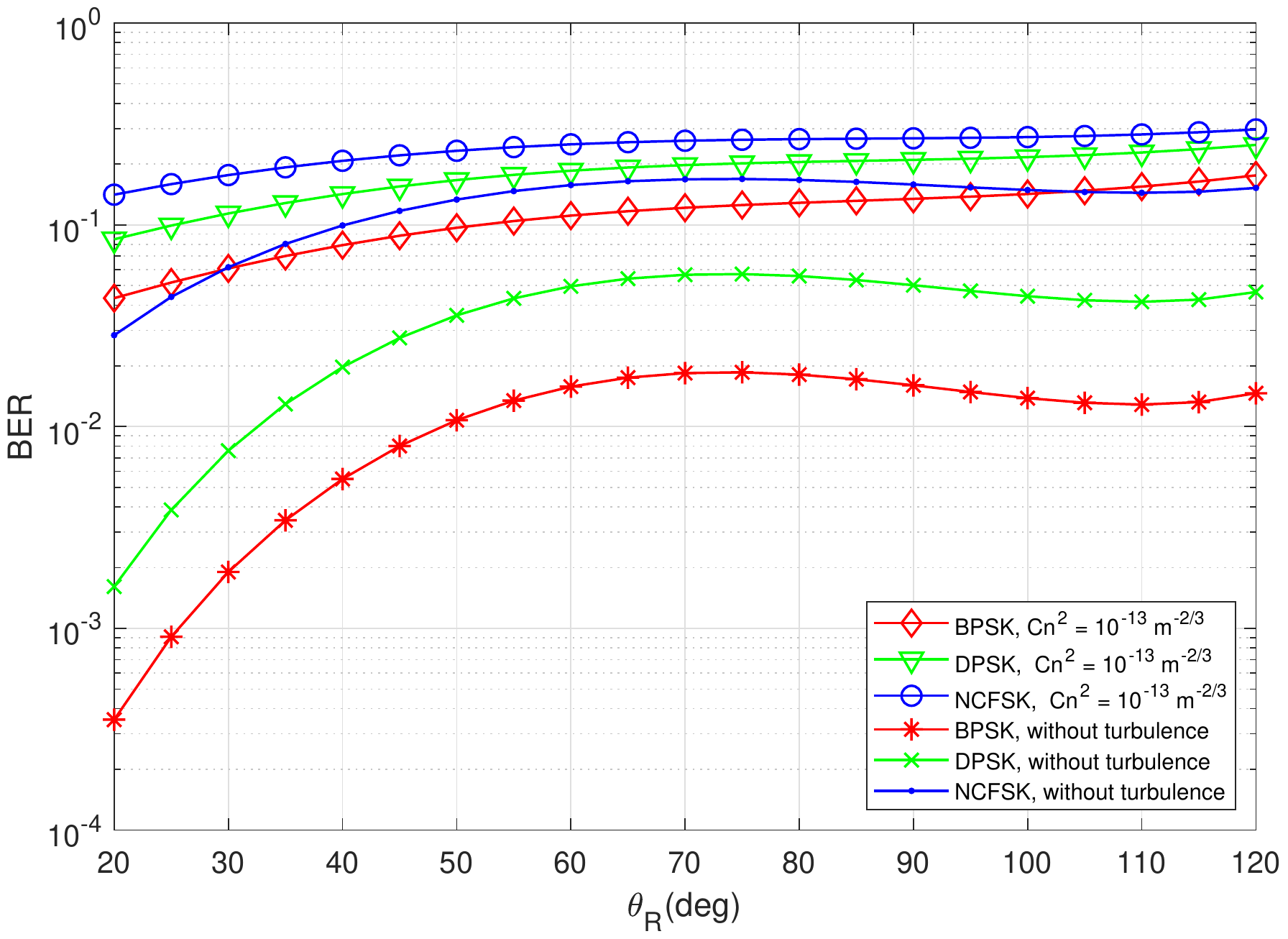}
		\caption{BERs of subcarrier BPSK, DPSK, and NCFSK modulated UV communication system for different receiver elevation angles with and without turbulence.}
		\label{Fig.5}
	\end{minipage}
\end{figure}

In Fig. 5, we consider the case of fixed transmitter elevation angles and  changable receiver elevation angles. The transceiver configurations are set as: $(\theta_T, \beta_T, \beta_R, r) = (30^{\circ}, 8~{\rm mrad}, 20^{\circ}, 200~{\rm m})$. In this case, the path loss is previously demonstrated to decrease in higher elevation angles \cite{shen2019modeling} due to larger common volume. From Fig. 5, we can see the same results for higher receiver elevation angles without turbulence. For example, when $\theta_R$ varies from $80^{\circ}$ to $115^{\circ}$, the BERs decreas from $1.72\times10^{-2}$ to $1.31\times10^{-2}$. However, when considering the influence of turbulence, we find that the BERs increase from $0.132$ to $0.148$.
 
\section{Conclusions}
In this work, we have derived a Meijer G form NLOS turbulence channel model. Based on this model, we have not only developped Meijer G form error rate expressions, but also developped approximate error rate results in finite series forms. Particularly, the approximate error rate results decompose the error rate into four parts corresponding to four turbulence parameters. Our asymptotic results indicate that the error rates mainly depend on the smaller turbulence parameters $\beta_1$ and$\beta_2$. We also find the influence of turbulence is strongest when the two LOS path through the same distance and the baseline range is longer than 300m assuming the same SNR, while the influence of turbulence is weakest when the two LOS path through the same distance and the baseline range is shorter than $200\,{\rm m}$. 


\bibliographystyle{IEEEtran}
\bibliography{IEEEabrv,ref}
\end{document}